\documentclass[a4paper,11pt]{article}
\usepackage{pos}
\usepackage{subcaption}
\usepackage{graphicx}
\usepackage{threeparttable}

\def\msun {{\mathrm{M}_\odot}}

\def\simless{\mathbin{\lower 3pt\hbox
     {$\rlap{\raise 5pt\hbox{$\char'074$}}\mathchar"7218$}}}   
\def\simmore{\mathbin{\lower 3pt\hbox
     {$\rlap{\raise 5pt\hbox{$\char'076$}}\mathchar"7218$}}}   

\title{Taxonomy of High Mass X-ray binaries from a historical perspective}
\ShortTitle{HMXBs from a historical perspective}

\author*[a,b]{Pablo Reig}

\affiliation[a]{Institute of Astrophysics, Foundation for Research and
Technology-Hellas, \\
71110, Heraklion, Crete, Greece}

\affiliation[b]{Physics Department, University of Crete,\\
70013, Heraklion, Crete, Greece}

\emailAdd{pau@physics.uoc.gr}

\abstract{
High-mass X-ray binaries serve  as important laboratories for studying
a broad range of fundamental astrophysical questions. These systems host two
distinct types of astrophysical objects at different stages of stellar
evolution, a massive donor star and a compact object. I will explore how
our understanding of these systems has evolved over the past 50 years, from the
dawn of the X-ray astronomy era to the present day. Along this historical
journey, I will introduce the various classes and sub-classes of high-mass
X-ray binaries and highlight their key observational characteristics.
}

\FullConference{87th Fujihara Seminar: The 50th Anniversary Workshop of the Disk Instability Model in Compact Binary Stars (DIM50TH2025)\\
22-26 September 2025\\
Tomakomai, Japan\\}

\begin{document}
\maketitle

\begin{figure}[htbp]
    \centering
    \includegraphics[width=0.99\textwidth]{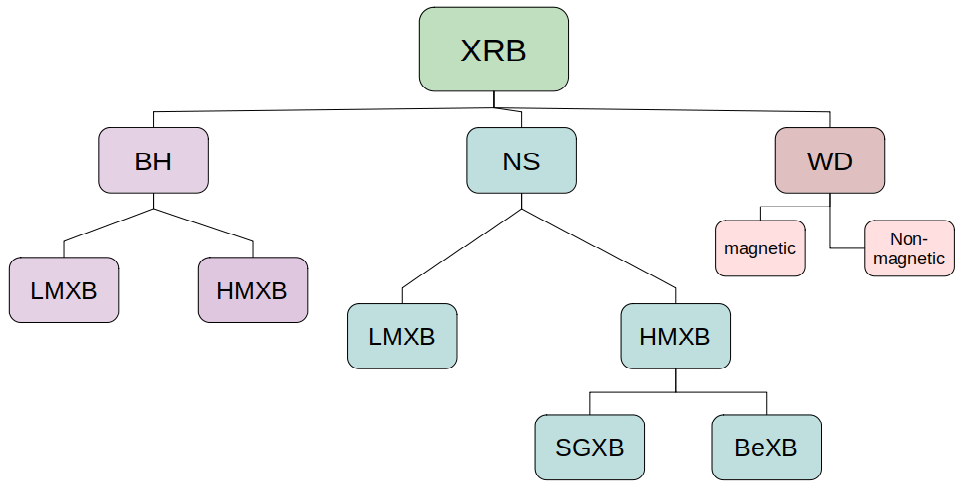}
    \caption{X-ray binaries classification.}
    \label{xrb-tree}
\end{figure}

\section{Introduction}

X-ray binaries consist of a compact object orbiting around a “normal” star.
They are "close" binary systems because there exists a transfer of mass
from the optical component to the compact object. By "normal" star, it is
understood that nuclear burning is still taking place in its interior, i.e.,
it is a star in the process of evolution. The compact object can be a black
hole, a neutron star, or a white dwarf; hence, we refer to them as black-hole
binaries (BHBs), neutron star binaries (NSBs), or cataclysmic variables (CVs),
respectively. The spectral type of the optical star defines whether the system
is a low-mass X-ray binary (LMXB) or a high-mass X-ray binary (HMXB). In
LMXBs, the spectral type of the optical companion is typically later than A,
while in HMXBs is an early-type B or late-type O star. In HMXBs, the mass
of the companion is typically larger than $\sim$ 8 $\msun$ and the class includes a
few BHBs (only one confirmed: Cyg X--1 and two strong candidates: Cyg X--3,
and SS 433) and about half of the NSBs. LMXBs, with the mass of the companion
below $\sim$2 $\msun$ include the majority of BHBs and half of the NSBs. CVs
contain low-mass companions and are normally considered a different class. The
luminosity class of the massive companion divides NS-HMXBs into supergiant
X-ray binaries (SGXBs) and Be/X-ray binaries (BeXBs). In SGXBs, the massive
star is a luminous evolved supergiant, while in BeXBs, it is a dwarf, subgiant,
or giant Be star. Figure~\ref{xrb-tree} depicts a tree diagram illustrating
the different types of X-ray binaries.  This work is focused on NS-HMXBs.

The importance of HMXBs stems from their role as
exceptional laboratories for studying a wide range of astrophysical phenomena.
Due to their young ages, HMXBs serve as tracers of recent star formation
activity \citep[e.g.,][]{grimm03,lutovinov05}. As a population, they provide
valuable insights into the properties of galaxies \citep[e.g.,][]{gilfanov04}.
The massive companion stars offer critical constraints on
stellar mass loss through winds, a key factor in stellar evolution models
\citep[e.g.,][]{gormaz23}. Furthermore, BeXBs enable
detailed investigations of the Be phenomenon, including the formation and
variability of circumstellar disks \citep{rivinius13, carciofi25}. 
The compact objects in HMXBs are powerful X-ray emitters, allowing us to probe
physical processes under extreme conditions: strong gravitational fields in BHBs
and NSBs, and intense magnetic fields in NSBs. These
systems facilitate long-term studies of accretion dynamics and accretion disk
physics \citep{mellah18,weng24}. Additionally, HMXBs allow for precise
measurements of neutron star masses, providing crucial constraints on the
nuclear equation of state \citep{quaintrell03,vandermeer07}. As close binary
systems, HMXBs represent a vital evolutionary stage in massive binary stars,
offering insights into the formation channels of neutron stars and black holes
\citep[e.g.,][]{vandenheuvel06}. Moreover, they are considered likely
progenitors of short gamma-ray bursts and sources of gravitational waves
\citep{kruckow18, roy21, petrosian24}.

In this work, we present a general overview of the taxonomy, historical
development, and observational properties of HMXBs, synthesizing key findings
from the literature.

\section{The 1960s: the beginning}

We start our journey back in the early 1960s when the first X-ray source of
extrasolar origin was discovered. This result was reported in a paper published
by Giacconi and collaborators in 1962: 

\begin{itemize}
\item "\emph{It is clear that the observed source does not coincide with any
    obvious scattering body belonging to our solar system}"  \citep{giacconi62}.
\end{itemize}

The source was detected relatively close to the galactic center in the Scorpio
region. Owing to the poor angular resolution of the detector, an association
with the galactic center could not be determined. However, new observations
performed soon after ruled out a physical connection with the galactic center
and confirmed the origin to be a point source in the Scorpio region:

\begin{itemize}

\item "\emph{A strong source was observed centered about the direction R.A. 16 h
10 m and Dec. $-18^{\circ}$ in the general proximity of v Scorpii. Assuming all
the emission to be concentrated at about 5 \AA, the computed flux is about
$1.5\times10^{-7}$ erg cm$^{-2}$ s$^{-1}$}" \citep{bowyer64}.

\item "\emph{We have observed two separate and intense sources of cosmic X-rays
in the region of the constellations Scorpius and Sagittarius during a rocket
experiment launched on August 28, 1964.}"  \citep{giacconi64}.

\end{itemize}

This source was named Scorpio X-1 (Sco X-1). A few years later, Centaurus X-3
(Cen X-3) was discovered, first as an X-ray source  \citep{chodil67} and then as
an X-ray pulsar:

\begin{itemize}
  
\item "\emph{In the 150-s exposure several peaks and valleys occur
with great regularity, immediately suggesting a periodic phenomenon... The
period of the fundamental frequency resulting from this fit is $4.832\pm 0.004$
s.}" \citep{giacconi71}. 

\end{itemize}

New sources were quickly discovered and by April 1968, about 30 X-ray sources
were known. Most of them were considered to belong to our Galaxy, but a few were
found at high galactic latitudes and were suggested to be associated with
external galaixies \citep{giacconi68}. A new window in the electromagnetic
spectrum had opened and the main debate at that time was to find an explanation
to the origin of the X rays:

\begin{itemize}
  
\item "\emph{An important question which must be answered about the recently
discovered stellar x-ray sources is the nature of the emission spectrum of these
sources.}" \citep{chodil65}

\item "\emph{No generally accepted model exist for any of the X-ray sources,
which is, in part, a reflection of the relative crudeness of the X-ray
measurements. The two physical mechanisms that seem to be capable of providing
the X-ray emission are the synchrotron radiation from ultrarelativistic
electrons gyrating in a magnetic field and the thermal radiation from an
extremely hot plasma.}" \citep{giacconi68}.

\end{itemize}

One process that was recognized from the very beginning as capable of powering X-ray
emission is accretion. As early as 1967, \citep{shklovsky67} proposed that an
accreting neutron star serves as the source of X-rays from Sco X-1:

\begin{itemize}
  
\item "\emph{If the identification of the optical object similar to an old nova
with the X-ray source is correct, then the natural and very efficient supply of
gas for such a accretion is a stream of gas, which flows from a secondary
component of a close binary system toward the primary component which is a
neutron star...The flux of gas in the stream is estimated as $10^{16}-10^{17}$
gm/sec ($\sim10^{-9} \, \msun$/year). When this gas falls on the neutron star the
production of energy per unit mass may amount to $\sim 10^{20}$ ergs/gm. Thus it
follows that the suggested modification of the mechanism of the accretion of gas
on the neutron star gives the possibility of explaining the power of X-ray
emission of the source Sco X-1.}" \citep{shklovsky67}.

\end{itemize}

The crucial words in the \citep{shklovsky67} extract are without a doubt {\it
close binary system}. For accretion to occur, an accreting object must be
present (in this case a neutron star), but there must be a source of matter.
Such a source arises naturally in a binary system.

\section{The 1970s: the binary model}

In the early 1970s, it was not yet conclusively established that Galactic X-ray
sources were binary systems, although the hypothesis was widely spread within
the astrophysical community.

\begin{itemize}
  
\item "\emph{At the time of writing (February-March 1972), the bandwagon,
    associated with the idea that many if not most of the powerful galactic
    X-ray sources are generated in binary star systems containing at least one
    exotic object together with gas, is gaining momentum.}" \citep{burbidge72}.
\end{itemize}



Three different types of systems were identified:

\begin{itemize} 
\item "\emph{The primary can either be a white dwarf, neutron star, or black
hole.}" \citep{burbidge72}.
\end{itemize}

Soon, the binary model consolidated:

\begin{itemize}  
\item "\emph{...reflects the preponderance of binaries
among all of the galactic X-ray sources. In fact, it can be argued that all the
X-ray sources are binaries.}" \citep{gursky75} 
\end{itemize}

People realized that that there were binaries with high and low-mass 
companions: 

\begin{itemize}
\item "\emph{One group of sources is associated with a very
particular kind of stars --- a late O or early B supergiant, a star which is
very massive and very luminous... Other X-ray sources are clearly not
associated with O-B stars, but rather with a much later spectral type stars ---
stars like the Sun in temperature, luminosity and mass.}" \citep{gursky75}
\end{itemize}

Among the HMXBs, two different types were identified depending on whether the
massive companion is an evolved star (i.e supergiant) or a star still in the
main sequence branch (BeXBs). Because of their brightness and persistent X-ray
emission, SGXBs were the first to be discovered.  They were initially thought to
represent the dominating population of HMXBs, whereas BeXBs were considered
atypical cases. Hence, the name classical or standard was given to SGXBs. BeXBs,
however, were the best candidates for transient X-ray behavior:

\begin{itemize} 
\item "\emph{We propose that sudden variations in the rate of
mass ejection from Be stars in the presence of a compact companion could produce
transient X-ray emission, which might recur over a period of years, like the
observed optical emission of Be stars. The companion star should be a neutron
star, rather than a white dwarf, because the spectrum of the transients is
rather hard (a = 1.0 for pulsating sources, and 2-4 for the others) and the
luminosity rather high  ($\sim 10^{37}$ erg s$^{-1}$). The total amount of
accreted matter required for an outburst is, for a neutron star, $\Delta M \sim 10^{23}$
g, which is only a few per cent of the amount of mass in the envelopes of Be
stars required for producing the observed emission lines.}" \citep{maraschi76}.
\end{itemize}

A key advance of this decade has been the establishment that the compact objects
in HMXBs are strongly magnetized neutron stars, and that these systems host
intense magnetic fields. These magnetic fields manifest in two primary ways:
through coherent X-ray pulsations and cyclotron resonance scattering features
(CRSF or simply cyclotron lines):

\begin{itemize} 
\item "\emph{An alternative and more likely picture for producing the pulsations requires
the funnelling of matter along the magnetic field lines to the surface of the
neutron star, where the released gravitational potential energy heats a small
area on the surface ($\sim$1 km$^2$) to extremely high temperatures ($\sim 10^8$ K). As the
neutron star rotates about its axis (which is presumably not aligned with the
magnetic axis), a distant observer will pass over a range of magnetic
coordinates and will thereby observe a corresponding X-ray pulse structure.}"
\citep{rappaport77}.
\end{itemize}

HMXBs also display cyclotron lines. The first cyclotron line was discovered in
Hercules X-1 (Her X-1), which although it is not a HMXB, it deserves the credit:

\begin{itemize}  
\item "\emph{We present further results of our Hercules X-1
balloon observation on 1976 May 3 which confirm the existence of a strong line
feature at $\sim$58 keV ... The most likely interpretation of this line is electron
cyclotron emission at the basic frequency from the hot polar plasma of the
rotating neutron star. The corresponding magnetic field strength is $5.3 \times
10^{12}$ gauss.}" \citep{trumper78}.
\end{itemize}

We now understand that virtually all HMXBs are X-ray pulsars, exhibiting pulse periods
ranging from a few seconds to several thousand seconds. Many of these systems
also display cyclotron lines in their X-ray spectra. Lines detected in the
energy range of 10--80 keV correspond to magnetic field strengths on the order
of $(1-10) \times 10^{12}$ G \citep{staubert19}.

\begin{figure}[t!]
    \centering
    \begin{tabular}{cc}
    \includegraphics[width=0.5\textwidth]{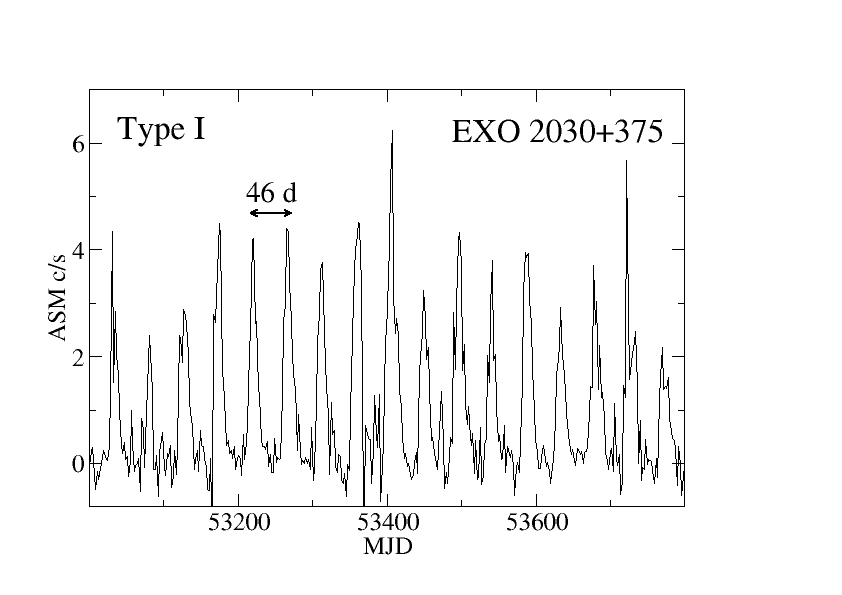} &
    \includegraphics[width=0.5\textwidth]{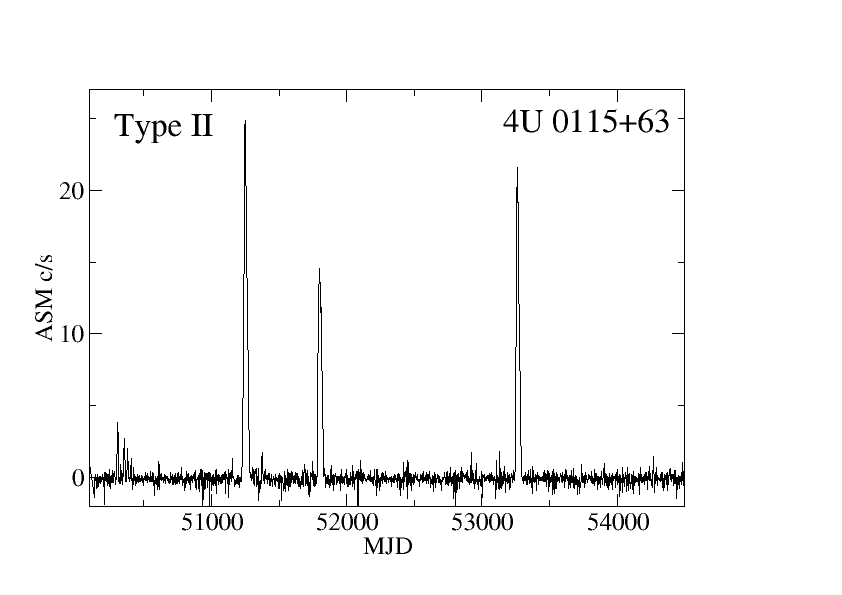} \\
    \end{tabular}
    \caption{Type I and type II outbursts in BeXBs.}
    \label{outbursts}
\end{figure}

\section{The 1980s:  mass transfer mechanisms}

At the beginning of the 1980s, the prevailing understanding was that that HMXBs
comprised two principal classes, namely,  SGXBs which are persistent sources
and BeXBs which are transient sources: 

\begin{itemize}  
\item "\emph{The BeXB are found to be systematically wider systems, with
lower-mass primaries, and with significantly transient behaviour than the
standard massive X-ray binaries such as Cen X-3 and SMC X-1.}"
\citep{rappaport82}.
\end{itemize}

In BeXBs, the optical companion is a Be star. The defining
observational characteristics of Be stars include emission lines --- most
notably in the Balmer series --- an infrared excess (i.e., an increase in flux
at longer wavelengths), and a small degree of linear polarization (typically
$\simless4$\%). These features are attributed to the presence of a circumstellar
disk orbiting the star around its equatorial plane. This disk forms from
material ejected from the stellar photosphere and evolves on timescales of
years: it can form, grow, and dissipate over time. When the disk is absent, the
emission lines revert to absorption, the infrared excess disappears, and no
polarization is detected. The long-term X-ray variability of BeXBs is described
by two types of transient activity or outbursts:

\begin{itemize}  

\item "\emph{Class I, periodic transient activity. ---- Transient outbursts that
recur periodically and have $L_X (\rm{max})/L_X (\rm{min}) \sim 100$... the
neutron star is typically in a moderately eccentric orbit and that the outbursts
occur close to the time of periastron passage. This suggests enhanced accretion
caused by the closer proximity of the neutron star to the Be star companion.}"

\item "\emph{Class II, irregular transient activity. --- Transient outbursts
that typically last several tens of days and involve an increase in luminosity
in excess of a factor of 100--1000. The timing of these outbursts is unrelated to
any underlying orbital period.}" \citep{stella86}.

\end{itemize}

\begin{figure}
    \centering
    \includegraphics[width=0.8\textwidth]{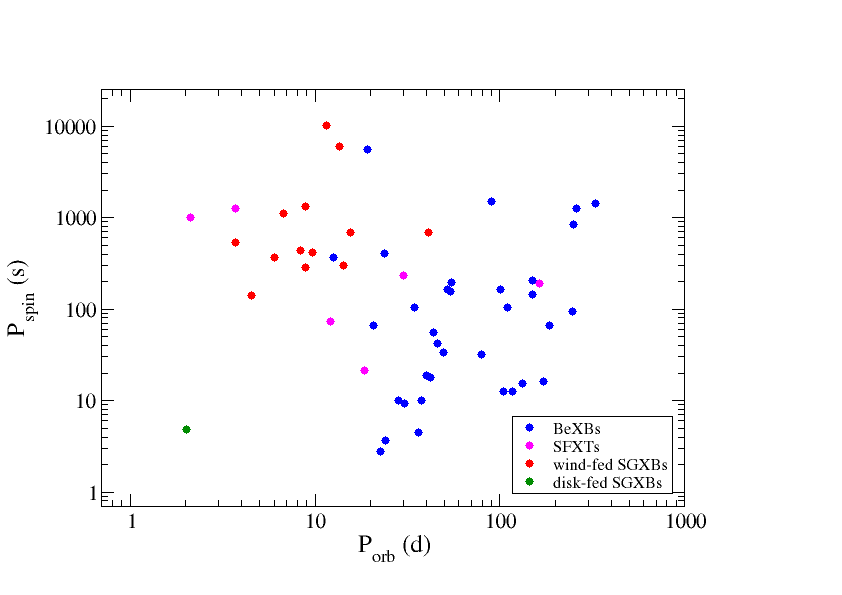}
    \caption{Updated version of the pulse period--orbital period diagram of Galactic HMXBs.}
    \label{psporb}
\end{figure}

Figure~\ref{outbursts} shows two typical examples of these type of outbursts. As more sources
were discovered, SGXBs were further divided into low-luminosity ($10^{36}$ erg
s$^{-1}$) and high-luminosity  ($10^{36}$ erg s$^{-1}$) sources. 

A major breakthrough in the understanding of HMXBs was the recognition that
the distinct subclasses correspond to fundamentally different modes of mass
transfer.

\begin{itemize}  

\item "\emph{A natural way of accounting for the separation of high-mass
pulsators into three groups is by invoking different mass-transfer processes for
the groups. These are, respectively, (i) -- accretion from the circumstellar
envelope of a Be star, (ii) -- accretion from the strong wind of an early-type
supergiant, and (iii) -- Roche lobe overflow}" \citep{corbet86}.
\end{itemize}

Interestingly,  these three subtypes of HMXBs occupy different regions in a
diagram where the pulse period is represented as a function of the orbital
period. 

\begin{itemize}  
\item "\emph{When segregated this way .... these classes appear to cluster in
different regions of the $P_{spin}$ versus $P_{orb}$ diagram.}" \citep{corbet86}. 
\end{itemize}

Figure~\ref{psporb} shows an updated version of this diagram. BeXBs are located in a
diagonal branch (blue circles), wind-fed SGXBs populate a horizontal branch in the upper left
part of the diagram (pink and red circles), and disk-fed SGXB the lower left
part (green circle). 

\begin{table*}
\centering
\caption{Observational properties of HMXBs as known by the end of the 1980s}
\label{hmxb-1980}
\begin{tabular}{|p{1.35cm}|p{1.7cm}|p{1.3cm}|p{1.85cm}|c|c|c|c|} 
\hline
Type	&Donor	&Compact	&Mass		&$P_{\rm orb}$	&$P_{\rm spin}$	&$L_X$	&Variability\\
	&star	&object		&transfer	&(days)		&(seconds)	&erg/s	& \\
\hline
BeXBs	&Be star \hspace{0.2cm}III-V & Neutron star	&Disk-fed (decretion$^*$)	&10--100	&10-100	&$10^{37}$	&Transient \\
\hline
SGXBs Low $L_X$	&O-B supergiant I-II	&Neutron star	&wind-fed	&4-10	&100-1000	&$10^{36}$	&Persistent  \\
\hline
SGXBs High $L_X$&O-B supergiant I-II	&Neutron star	&disk-fed (accretion)&1-4	&1-10	&$10^{38}$	&Persistent  \\
\hline
\end{tabular}
\begin{tablenotes}
\item $^*$ {\small The term "decretion disk" in the context of Be stars appears to
have been formally introduced in the early 1990s.}
\end{tablenotes}
\end{table*}

\section{The 1990s: the discovery of persistent BeXBs}

Our knowledge of HMXBs by the end of the 1980s is summarized in
Table~\ref{hmxb-1980}. In particular, BeXBs appeared to be transient and
moderately eccentric systems:

\begin{itemize}  
\item "\emph{The most notable feature of the Be X-ray binaries is that they are very often
bright transient sources... Pulsations are usually
detected, many (but not all) with periods around a few seconds. Doppler
variations in the pulse period give orbital periods of a few tens of days, with
a moderate eccentricity ($e \sim0.3$).}" \citep{white89}
\end{itemize}

\begin{figure}
    \centering
    \includegraphics[width=0.8\textwidth]{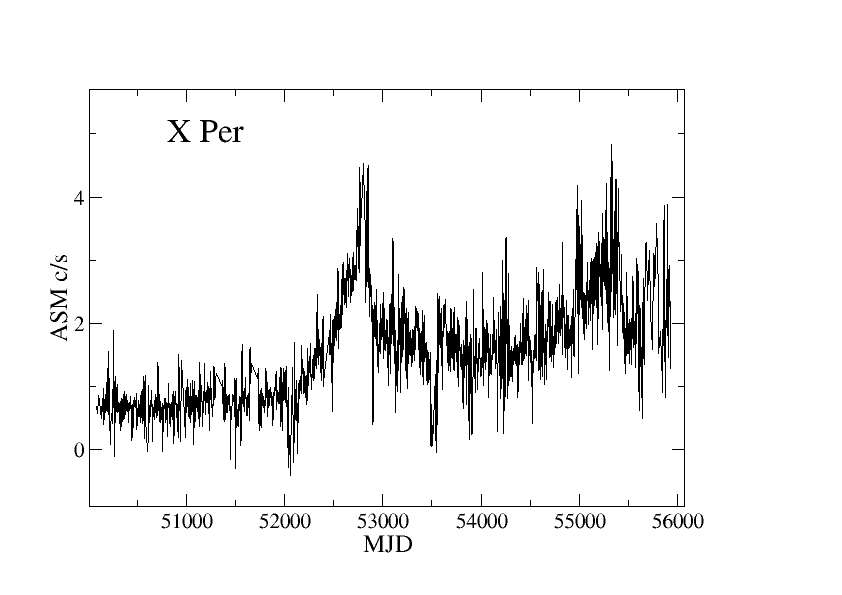}
    \caption{Long-term X-ray variability of X Per.}
    \label{xper}
\end{figure}

Thanks to the improvement of the sensitivity of the X-ray detectors, {\it
ROSAT} and {\it RXTE} uncovered a new type of BeXBs characterized by persistent
X-ray emission:

\begin{itemize}  
\item "\emph{It therefore seems that we now have a growing subclass of BeXRBs, characterized
by persistent, low-luminosity X-ray emission and slowly rotating pulsars. A
possible model fo these systems is that of a neutron star orbiting a Be star in
a relatively wide orbit, accreting material from only the low-density outer
regions of the circumstellar envelope.}" \cite{reig99}.
\end{itemize}

\noindent with the following properties: (i) long pulse periods, (ii)
persistent, low-luminosity ($10^{34}-10^{35}$ erg s$^{-1}$) X-ray emission; (iii) low cut-off
energy, $E_{\rm cut}$ < 4--5 keV ; (iv) absent or very weak iron line at 6.4 keV,
indicative of only small amounts of material in the vicinity of the neutron
star; (v) low X-ray variability: flat light curves with rare and unpredictable
increases in flux by a factor of $<10$;  (vi) no dependence of the X-ray spectrum
on intensity. The prototype of persistent BeXBs is X Persei
(X Per). Figure~\ref{xper} shows its long-term X-ray light curve.

\section{The 2000s: the discovery of new subtypes of HMXBs}

The 2000s represent a golden decade in the study of HMXBs in terms of the number
of new sources detected for the first time and the discovery of low eccentricity
BeXBs and transient SGXBs. Prior to the launch of the {\it INTEGRAL} mission in
October 2002, the population of BeXBs was increasing rapidly, whereas the number
of SGXBs had reached a steady level. Throughout the 1980s and 1990s, the
discovery rate of new systems was approximately four to one in favor of BeXBs.
This disparity can be attributed to the sensitivity of the detectors and the
different nature of the X-ray emission: since SGXBs are persistent emitters,
further discoveries primarily resulted from improvements in the sensitivity of
X-ray detectors aboard space missions.  Although BeXBs also benefited from these
technological advancements, their detection additionally depended on the
episodic activation of their transient outbursts.

As pointed out above, the orbit in the vast majority of BeXBs was rather
eccentric ($e>0.3$). This eccentricity results from a large kick during the
supernova explosion that leads to the formation of the NS. Thus, it was
surprising to discover systems with nearly circular orbits:

\begin{itemize}  
\item "\emph{...a new observed class of HMXBs with orbits that are distinguished by
relatively long periods ($P_{\rm orb} \sim 30-250$ days) and low eccentricities
($e< 0.2$). Members of this new class of HMXBs contain NSs that almost certainly received a
fairly small kick ($< 50$ km s$^{-1}$) at the time of formation.}" \citep{pfahl02}.
\end{itemize}

The explanation for such low orbital eccentricities lies in the dependence of
the natal kick imparted to the neutron star on the rotation rate of its
progenitor's core following mass transfer. According to \citep{pfahl02}, rapidly
rotating pre-collapse cores produce NSs with small kicks, whereas slowly
rotating cores result in larger kicks. If the progenitor's envelope is stripped
before deep convection develops, the exposed core retains rapid rotation;
conversely, if mass transfer occurs after significant evolution, magnetic
torques between the convective envelope and the core can spin down the core to
very low rotation rates.

Another relevant result of this decade was the identification of a new subclass of
BeXBs was proposed. Although the prototype of this class, $\gamma$  Cas, had
long been well studied, particularly in the ultraviolet and optical bands,
it also exhibits X-rays emission. The realization that $\gamma$ Cas was not an isolated
case, but instead representative of an entire new subclass, emerged when several
newly discovered X-ray sources were found to share key properties with $\gamma$
Cas.

\begin{itemize}  
\item "\emph{It is now clear that the ... $\gamma$ Cas puzzle is no longer an unique case, ...
They constitute a new and well-defined class of X-ray emitters that is
characterised by: [1] a hard X-ray emission that is very likely thermal 
and dominated by a component with $7 < \rm{kT (keV)} < 13$, [2] a variable behaviour,
[3] a moderate 0.2--12 keV luminosity ($32 < \log L_X (\rm{erg s}^{-1} ) < 33$), and from
their optical properties, by [4] a large and probably stable circumstellar
disc.}" \citep{lopes06}.
\end{itemize}


These systems were designated $\gamma$ Cas analogues. While their optical
properties are consistent with those of BeXBs, their X-ray properties differ significantly. The
optical companion in both BeXBs and $\gamma$ Cas analogues are late-type O or
early-type B III-V stars. They both exhibit long-term optical spectral
variability, especially in the H$\alpha$ equivalent width, that can be
attributed to the Be star's circumstellar disk.  

In contrast, their X-ray emission is predominantly thermal. Indeed, about 90\%
of the X-ray spectrum of  $\gamma$ Cas in the energy range 1--100 keV can be
fitted with a purely thermal plasma component at $\sim15$ keV (see Fig. 5 in
\citep{shrader15}). The X-ray spectra of BeXBs is clearly non-thermal,
characterized by a power law and a high-energy cut off, indicative of
Comptonization. Moreover, the X-ray luminosity of $\gamma$ Cas analogues is
significantly lower than BeXBs. Typical luminosities in $\gamma$ Cas binaries
fall in the range $10^{32}-10^{33}$ erg s$^{-1}$, compared to $10^{34}-10^{35}$
erg s$^{-1}$ for persistent BeXBs, $10^{36}-10^{37}$ erg s$^{-1}$ during type I
outbursts, and $10^{37}-10^{38}$ erg s$^{-1}$ during type II outbursts. Another
difference is that among $\gamma$ Cas binaries, no X-ray pulsars have been
found.  

Three different scenarios have been put forward to explain the X-ray emission
from $\gamma$ Cas analogues: the magnetic model states that X-rays a produced
from the interaction of the small-scale magnetic field of the Be star and the
magnetic field in the disk. Rapid stellar rotation drives magnetic reconnection
events, accelerating electrons that subsequently produce X-ray emission
\citep{smith16}; the second scenario involves accretion onto a white dwarf or a
NS in the propeller regime \citep{postnov17, gies23}; the third possibility is that
X-rays arise from shocks produced by collision between the disk material and the
wind of a hot subdwarf or He-rich companion star \citep{langer20}. 

The 2000s also witnessed the discovery of transient SGXBs. Up until the launch
of {\it INTEGRAL}, our understanding was that SGXBs were persistent X-ray
sources. However, {\it INTEGRAL} uncovered a new class of X-ray sources with
supergiant companions transient X-ray emission:

\begin{itemize}  
\item "\emph{These outbursts are very short (lasting
from $\sim$3 to $\sim$ 8 hours) and present very sharp rises, reaching the peak of
the flare in < 1 h.... We therefore propose that all these objects form a class
of HMXBs which we call Supergiant Fast X-ray Transients (SFXTs), because of the
fast outbursts and super giant companions. They differ from classical wind-fed
SGXBs, whose X-ray luminosity is variable but always detectable around $L_X
\sim10^{36}$ erg s$^{-1}$.}" \citep{negueruela06}.
\end{itemize}

The main observational characteristics of SFXTs are: (i) short outbursts, lasting
hours to days compared to weeks or months in transient BeXBs; (ii) often show an
asymmetric profile with a fast rise, followed by an exponential decay: the  rise
time is typically minutes to hours, whereas the decay times is hours to days.
Some systems exhibit a multiple peaks profile within a single outburst; (iii) 
low duty cycle:  SFXTs spend less than 5\% of their time in high-luminosity
flaring states; (iv) high dynamic range: the ratio between maximum and minimum
X-ray luminosity is $\simmore 100$, although some extreme cases display up
four orders of magnitude difference in luminosity during flares; low
time-averaged luminosity ($L_X \simless 10^{34}$ erg s$^{-1}$), often
undetectable in quiescence and only visible during bright flaring events.

SFXTs tend to occupy the same region in the $P_{\rm spin}-P_{\rm orb}$ diagram
as the wind-fed SGXBs. However, as can be seen in Fig.~\ref{psporb}, some SFXTs
lie at intermediate position between BeXBs and wind-fed SGXBs or even fall on
top of the BeXBs region (IGR J11215-5952). Moreover, the long-term X-ray
variability of IGR J17391--3021 is reminicent of BeXBs type I outbursts
\citep[see Fig.2 in][]{drave10}. These similarities open the possibility that
SFXTs may be the descendants of BeXBs \citep{liu11}.

\section{The 2010s: new discoveries}

Thanks to the new generation of $\gamma$-ray detectors, especially from the
mid-2000s onwards, a new class of HMXBs has emerged. They are called
Be/$\gamma$-ray binaries because most of their radiated power is emitted beyond
1 MeV:

\begin{itemize}  
\item "\emph{The five binaries detected in VHE gamma rays form a new class of systems,
gamma-ray binaries, distinguished by emitting most powerfully beyond 1 MeV and
by having a O or Be massive star companion.}" \citep{dubus13}.
\end{itemize}

The main observational properties of Be/$\gamma$-ray binaries are: (i) modest
X-rays but strong $\gamma$-ray emission ($> 1$ MeV); (ii) massive stellar
companion:  the optical counterpart is either a luminous O-type star (LS 5039),
a Be star (LS I +61 303, PSR B1259-63, HESS J0632+057) or a likely Wolf-Rayet
star (Cygnus X-3); (iii) radio emission, in contrast to BeXBs which are radio
quiet sources; (iv) orbital modulation at all wavelengths, from radio to
$\gamma$ rays. The NS in these systems is supposed to be recently formed, hence it is
rotating very fast and ejecting a strong pulsar wind:

\begin{itemize}  
\item "\emph{What makes them so bright in gamma rays? The prevailing idea is that
gamma-ray  non-thermal emission is due to particles accelerated at the shock
between the wind of the massive star and the wind of a pulsar.  Hence, gamma-ray
emission is ultimately powered by the spin-down of a rotating neutron star with
a strong magnetic field $\sim 10^{11}-10^{13}$ G. They are pulsar wind
nebulae in a binary environment.}" \citep{dubus13}.
\end{itemize}

Another intrguing unresolved observational puzzle of this decade is the lack of
Be star binaries with BHs. By 2009, not a single BeXB with a black hole as the
compact object had been discovered. However,  there seemed to be no apparent
mechanism that would prevent the formation or detection of Be stars with BHs.  
According to evolutionary scenarios, the ratio of binaries with NSs to the ones
with BHs was estimated to be as high as $\sim50$. In this case, the non
detection of Be-BH systems would be simply a question of small number
statistics:

\begin{itemize}   
\item "\emph{...we know 64 BeXRBs in the Galaxy, but only 42
of these systems are known to host an NS. None of the observed BeXRBs hosts a
BH....We predict that both populations of BeXRBs should exist in the Galaxy...If
we use the preferred evolutionary models ($F_{\rm NStoBH} \sim 30-50$...we
predict that in the observed sample of BeXRBs of 42 systems with NSs, one should
expect only $\sim0-2$ systems with BHs. It is quite possible that none are yet
observed (small statistics).}" \citep{belczynski09}.
\end{itemize}

However, the number of BeXBs was increasing ($\sim80$ by mid 2010s) and yet no Be-BH binaries was being
found. Finally the discovery of a Be-BH binary was reported:

\begin{itemize}   
\item "\emph{MWC 656 (=HD 215227)... Our adopted B1.5--B2 III classification implies a
mass of 10--16 solar masses for the Be star, and hence a companion star of
3.8--6.9 solar masses}" \citep{casares14}.
\end{itemize}

Another BH-HMXB is HD96670, an O8 main-sequence star with a massive secondary
component.

\begin{itemize}   
\item "\emph{HD96670: we model the optical light curve and radial velocity curve
simultaneously ... and find a best-fit mass of $M_1 = 22.7^{+5.2}_{-3.6}$ solar masses
for the primary, and $M_2 = 6.2^{+0.9}_{-0.7}$ solar masses for the secondary. An
object of this mass is consistent with either a B-type star, or a black hole.
Given that we see no absorption lines from the secondary, in combination with an
observed hard power-law X-ray spectrum ... that may have been produced by wind
accretion onto the secondary, we conclude that the secondary is most likely a
black hole.}" \citep{gomez21}.
\end{itemize}

At present, these two systems MWC 656 and HD96670 remain as BH-HMXBs candidates
and more observations are needed to better sample the radial velocity curve and
to understand the nature of the secondary component. We note that HD96670 is not
a OBe star: there is no evidence for a decretion disk. On the contrary, there is
some evidence of the presence of a third star in this system. The case of MWC 656
result has been questioned \citep{janssens23}.


\begin{figure}[t!]
    \centering
    \includegraphics[width=0.99\textwidth]{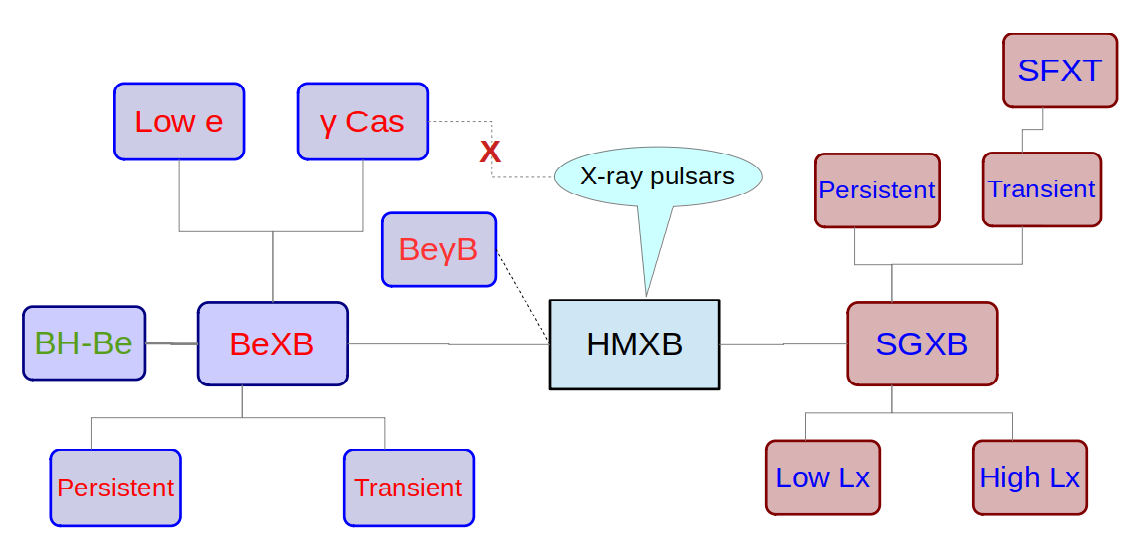}
    \caption{The "zoo" of HMXBs.}
    \label{zoo}
\end{figure}

Figure~\ref{zoo} summarizes the various classes of HMXBs, as they are known
today.

\section{Conclusions}

HMXBs are diverse and fascinating systems offering insight into numerous
astrophysical problems, including accretion physics, compact object properties,
binary evolution, stellar winds and magnetic interactions to mention just a few.
Current estimates set the number of HMXBs in more than 150 systems: about half
of them are BeXBs, 34\% of SGXBs, and 16\% of uncertain classification. Although
the most numerous subgroup are BeXBs, the growth in the number of confirmed
SGXBs during the last 15 years has been extraordinary, going from 16 systems in
2006 to 52 in 2023 \citep{fortin23}.


\end{document}